\documentclass[useAMS,usenatbib]{mn2e}
\usepackage{epsfig}
\newcommand{\mnras}{\,{\rm MNRAS}}
\newcommand{\apj}{\,{\rm ApJ}}
\newcommand{\apjl}{\,{\rm ApJL}}
\newcommand{\apjs}{\,{\rm ApJS}}
\newcommand{\aap}{\,{\rm A\&Ap}}

\newcommand{\aj}{\,{\rm AJ}}
\newcommand{\prd}{\,{\rm PRvd}}

\newcommand{\ssst}{\scriptscriptstyle}

\newcommand{\etal}{et al.}

\newcommand{\yr}{\,{\rm yr}}    
\newcommand{\cm}{\,{\rm cm}}

\newcommand{\no}{n_{\ssst 0}}   \newcommand{\ti}{t_{\rm i}}
\newcommand{\Rc}{R_{\rm c}}     \newcommand{\Rs}{R_{\rm s}}
    
\newcommand{\ESNR}{E_{\ssst\rm SNR}}
        \newcommand{\mH}{m_{\ssst\rm H}}

\newcommand{\gray}{$\gamma$-ray}    \newcommand{\grays}{$\gamma$-rays}

\newcommand{\Spitzer}{{\sl Spitzer}}
  \newcommand{\Fermi}{{\sl Fermi}}

\title[GeV spectra break for interacting SNRs]
{\grays\ from molecular clouds illuminated by accumulated diffusive protons. II: interacting supernova remnants}

\author[H. Li et al.]
{Hui Li$^{1}$
~and Yang Chen$^{1,2}$\thanks{E-mail: ygchen@nju.edu.cn}%\footnotemark[1]\thanks{This file has been amended to}
\\
$^{1}$Department of Astronomy, Nanjing University, Nanjing 210093, P.\ R.\ China\\
$^{2}$Key Laboratory of Modern Astronomy and Astrophysics, Nanjing University, Ministry of Education, Nanjing 210093, China\\
}

\begin{document}

\date{}

\pagerange{\pageref{firstpage}--\pageref{lastpage}} \pubyear{2008}

\maketitle

\label{firstpage}

\begin{abstract}
Recent observations reveal that spectral breaks at $\sim$GeV are commonly present in Galactic \gray\ supernova remnants (SNRs) interacting with molecular clouds and that most of them have a spectral ($E^2dF/dE$) ``platform'' extended from the break to lower energies. In paper~I (Li \& Chen 2010), we developed an accumulative diffusion model by considering an accumulation of the diffusive protons escaping from the shock front throughout the history of the SNR expansion. In this paper, we improve the model by incorporating finite-volume of MCs, demonstrate the model dependence on particle diffusion parameters and cloud size, and apply it to nine interacting SNRs (W28, W41, W44, W49B, W51C, Cygnus Loop, IC443, CTB 37A, and G349.7+0.2). This refined model naturally explains the GeV spectral breaks and, especially, the ``platform''s, together with available TeV data. We find that the index of the diffusion coefficient $\delta$ is in the range of 0.5-0.7, similar to the galactic averaged value, and the diffusion coefficient for cosmic rays around the SNRs is essentially two orders of magnitude lower than the Galactic average ($\chi\sim$0.01), which is a good indication for the suppression of cosmic ray diffusion near SNRs.
\end{abstract}


\section{Introduction}\label{sec:intro}

Supernova remnants (SNRs) are commonly believed to be one of the most important acceleration sites for the cosmic rays (CRs) below the ``knee" in our Galaxy. Diffusive shock acceleration (DSA) is the prevailing acceleration mechanism, which can naturally gain a power-law population of relativistic electrons/protons (Blandford \& Eichler 1987). Although multi-wavelength analysis has been intensively made for the emission of SNRs, it is still in hot debate whether the \grays\ from them are of hadronic or leptonic origin. Recently, \gray\ telescopes of new generation, such as \Fermi-LAT and H.E.S.S., provided more and more clues that SNRs interacting with molecular clouds (MCs) may emit \grays\ arisen from $\pi^0$ decay via proton-proton collision (e.g., Abdo \etal\ 2009, 2010a,b,c,d; Aharonian \etal\ 2008a,b).

Interacting SNRs, distinguished by several kinds of evidence such as OH masers, molecular line broadening, etc.(see Jiang \etal\ 2009 and references therein), represent a promising class of \gray\ sources.
%Excellent correlation exists between the SNR masers and a subset of GeV and TeV sources in the field of SNRs (Hewitt \etal\ 2009).
It is seen that some of the SNRs exhibit a spectral break at $\sim$GeV, such as W28, W44, W51C, IC443, etc. Most of them have a spectral ($E^2dF/dE$) ``platform'' extended from the break to low energies.  By coincidence, the \gray\ emitting SNRs which harbor OH masers (Hewitt \etal\ 2009) all have GeV breaks. The spectral break can be explained in the hadronic scenario either by different acceleration effects in the SNR-MC system, such as Alfv\'en wave evanescence in the weakly ionized dense gas (Malkov, Diamond \& Sagdeev 2010), reacceleration model for crushed clouds (Uchiyama \etal\ 2010), and two-step acceleration model in the reflected shocks (Inoue, Yamazaki \& Inutsuka 2010), time-dependent two-zone model (Tang \etal\ 2011), or by diffusion effects of CRs escaping from SNR shock (Li \& Chen 2010, hereafter Paper I; Ohira, Murase, \& Yamazaki 2011).

In Paper I, we developed an accumulative diffusion model by considering the accumulation of the diffusive protons escaping from the shock front throughout the history of the SNR expansion; the power-law distribution is assumed for the escaping protons and the spectrum of the diffusive protons at any point near the SNR is obtained. This model is used to explain the GeV break and GeV/TeV flux discrepancy for the four sources around SNR W28.
Comparing with the point source injection (Aharonian \& Atoyan 1996; Gabici et al.\ 2009), our model is more physical when applied to the finite-size accelerator, such as SNRs, and small distance between the accelerator and MC is allowed.\footnote{Actually, our model in Paper I has been tested for the case that the SNR has a very small radius (e.g., for a very high ambient density), which can be approximated as a point-like source, and reproduced the results of the classical point-like continuous injection case using the same model parameters.} Almost meantime, Ohira \etal\ (2011) have also developed a diffusion model by considering finite-size MCs interacted by the CRs escaping from the SNRs which are imbedded in MCs and used the model to fit the \gray\ spectra of four SNRs (W28, W44, W51C, and IC443). In their model, the time evolution of maximum particle energy is considered and mono-momentum is assumed for the escaping particles and the spectral shape of the diffusive protons is analytically derived.

In this paper, we improve the accumulative diffusion model by incorporating finite-volume of MCs (Sec.~\ref{sec:model}) and apply it to a series of interacting SNRs to reproduce the GeV-TeV spectra as well as the ``platform" below the break at around GeV (Sec.~\ref{sec:app}).

%We find that the platform before the break energy is linked with the length-scale of the MCs and it is the great touchstone for the interacting system.

\section{Accumulative diffusion Model with finite-volume MCs}\label{sec:model}
\subsection{Model description}\label{sec:MD}

%When MCs in the vicinity of SNRs are located at distances ($R_{\rm c}$) measured from the centre of the SNR comparable to the radius ($R_{\rm SNR}$) of the SNR itself, i.e., so close to the SNR blast shock that clouds are overtaken by the expanding shock, the diffusion distance for protons injected at various times, $\ti$, and various shock radii, $\Rs$, is no longer constant and the assumption of point-like source injection should be modified. Li \& Chen (2010) have enlarged the model setup of the previous section {\bf (?)} considering that cosmic-rays impacting into the MC are an accumulated collection of the diffusive protons escaping from different $\Rs$ as the SNR expands. In this approach, the distance between the SNR and MC does not need to be larger than the sizes of both objects.
In Paper I, the distribution function, at an arbitrary point (at radius $\Rc$), of the energetic protons that escape from the spherical shock front (at radius $\Rs$) throughout the history of the SNR expansion is given by
\begin{eqnarray}\label{eq:f_cum}
&f_{\rm cum}(E_{\rm p},\Rc,t_{\rm age})= \int_0^{t_{\rm age}}\int^{2\pi}_0\int^{\pi}_0 f(E_{\rm p},R(\Rc,\ti,\theta,\phi),t_{\rm dif})\nonumber\\
 &\Rs^2(\ti)\sin\theta\,d\theta\,d\phi\,d\ti,
\end{eqnarray}
where $t_{\rm age}$ is the age of the SNR, $t_i$ the time at which the proton escapes from SNR surface, $t_{\rm dif}$ the diffusion time after escape ($t_{\rm dif}=t_{\rm age} - t_i$), and $f(E_{\rm p},R(\Rc,\ti,\theta,\phi),t_{\rm dif})$ the distribution function, at point C, of the energetic protons that escape from unit area at an arbitrary source point S on the spherical shock front surface (see Paper I for relevant notations). The diffusion coefficient is assumed to be in the form of $D(E_{\rm p})=10^{28}\chi(E_{\rm p}/10\rm GeV)^\delta \rm cm^2\rm s^{-1}$, where $\chi$ is the correction factor of slow diffusion around the SNR (Fujita \etal\ 2009) and $\delta$ is the energy dependent index of diffusion coefficient. The distance between the SNR and MC does not need to be larger than the size of SNR, namely condition $\Rc\gg\Rs$ is not required as opposed to the point source injection models.

Following the treatment of the dynamical evolution of SNR in Paper I, %for the dynamical evolution of radius of the SNR expanding in the interstellar (intercloud) medium of density $\rho_0=1.4\mH \no$,
we use the Sedov-Taylor law $R_{\rm s}=(2.026 \ESNR t^2/\rho_{\ssst 0})^{1/5}$ for the adiabatic phase, where $\rho_{\ssst 0}=1.4\mH \no$ is the density of the interstellar (intercloud) medium, and $R_{\rm s}=(147\epsilon \ESNR R_{\rm t}^2t^2/4\pi\rho_{\ssst 0})^{1/7}$ for the radiative phase, where $R_{\rm t}$ is the transition radius from the Sedov phase to
the radiative phase, $\ESNR$ is the supernova explosion energy, and $\epsilon$ is a factor equal to 0.24 (Blinnikov \etal\ 1982; see also Lozinskaya 1992).

In Paper I, we assume that all the MC mass is concentrated in a point. Here we take the finite-volume of cloud into account and in order to describe the problem with spherical coordinates, we approximate it as a truncated cone which subtends solid angle $\Omega$ at the SNR center and has a thickness $\Delta \Rc$. By averaging the distribution function $f_{\rm cum}$ (Eq.~(\ref{eq:f_cum})) over the finite-volume, we obtain the mean distribution of the energetic protons in the MC at $t_{\rm age}$ as a function of $E_{\rm p}$
\begin{eqnarray}
F_{\rm ave}(E_{\rm p},t_{\rm age}) &=&\int_{\Rc-\Delta \Rc/2}^{\Rc+\Delta \Rc/2}r^2 dr
\int_0^{t_{\rm age}}\int^{2\pi}_0\int^{\pi}_0\nonumber \\
&&f(E_{\rm p},R(r,\ti,\theta,\phi),t_{\rm dif}) \Rs^2\sin\theta\,d\theta\,d\phi\,d\ti\nonumber\\
& & \bigg/\int_{\Rc-\Delta \Rc/2}^{\Rc+\Delta \Rc/2}r^2 dr.\label{eq:finite-volume}
\end{eqnarray}
%\begin{equation}
%F_{\rm tot}(E_{\rm p},t_{\rm age}) =\int_{\Rc-\Delta \Rc/2}^{\Rc+\Delta \Rc/2}r^2 dr
%\int_0^{t_{\rm age}}\int^{2\pi}_0\int^{\pi}_0
%f(E_{\rm p},R(r,\ti,\theta,\phi),t_{\rm dif}) \Rs^2\sin\theta\,d\theta\,d\phi\,d\ti% \bigg/\int_{\Rc-\Delta \Rc/2}^{\Rc+\Delta \Rc/2}r^2 dr.
%\end{equation}
In particular, for $\Omega=4\pi$, we have the case that the SNR is enclosed by a molecular shell (as also assumed in Ohira \etal\ 2011).

In the calculation of the $\gamma$-rays from the nearby MC (of mass $M_{\rm c}$) due to p-p interaction, we use the analytic photon emissivity $dN_{\gamma}/dE_{\gamma}$ developed by Kelner et al.\ (2006; also see Paper I). The gained photon energy is generally $10\%$ of the energetic proton energy (e.g.\ Katz \& Waxman 2008).

\subsection{Model Performance}\label{sec:MP}

%As mentioned in Section~\ref{sec:MD}, the model we have used here is slightly different from the model in paper I by considering the finite-size of the target MCs.
We now explore the model parameters and show how the spectra of escaping protons depends on the parameters. In the model described above, for a certain SNR, the model parameters are of two sorts: particle diffusion ($p$, $\delta$, $\chi$) and MCs ($\Rc$, $\Delta R$).
%Recent study shows that the correction factor of slow diffusion around the SNR, $\chi$, is approximated as $\sim0.1$ and we adopt this value throughout this paper (Fujita \etal\ 2009).

Figure~1 shows the dependence of the spectral shape of the energetic protons in the finite cloud volume on each parameter with other parameters fixed. Here we assume a middle-aged SNR with $R_{\rm SNR}=20$pc, $t=27000\yr$, $\no=1\cm^{-3}$, and $\ESNR=10^{51}$erg and a proton illuminated MC at $\Rc=25$pc.

%3333Impacts of finite-volume of MCs $\Delta R$
The spectral dependence on the MC thickness $\Delta \Rc$ is presented in Figure~1{\em a},
where we adopt $\chi=0.01$, $p=2.2$, and $\delta=0.5$. We plot three curves for different thicknesses: $\Delta \Rc\rightarrow0$, $\Delta\Rc=8$~pc, and $\Delta\Rc=10$~pc. In the $\Delta \Rc\rightarrow0$ case, the cloud volume is not considered, similar to the treatment in Paper I. We see that there are spectral breaks in the diffusive protons and find that, with the thickness increased, the slope change around the break becomes gentle and a ``platform"-like pattern appears and gets broadened. This pattern is formed by the combination of particle distribution at different diffusion distances (within the cloud thickness), as the spectral peak shifts to higher energies with the increase of distance. %This platform will also exhibit in the \gray\ spectra.
The $\Delta\Rc=10$~pc case represents the situation that the MC is in contact with the SNR shock surface ($\Rc=\Rs+\Delta\Rc/2$; such case is called `contact' case hereafter), in which the proton spectrum does not show a low-energy cutoff/turn-over since some of the low-energy protons can diffuse into the cloud.

%Fig ... shows the spectral dependence on the escaping protons energy index $p$

The spectral dependence on the escaping protons energy index $p$ is given in Figure~1{\em b},
with parameters $\chi=0.01$, $\Delta R=15$pc, and $\delta=0.5$ used. %(relative large value in order to see clearly of the platform)
Three values of $p$ are used for comparison: $p=1.8$ is the particle spectral index predicted by the DSA mechanism considering non-linear effects, $p=2$ is the typical index of Fermi-type accelerated particles, and $p=2.2$ is the source spectral index of the Galactic CRs derived from the diffusion model GALPROP %derived from the Galaxy diffusion model
(Strong \etal\ 2007). %This parameter can influence both the slope of the platform and the spectral index after the break.
For the `contact' case ($\Delta\Rc=10$~pc), %the slope at energies below and above the break will be smaller when $p$ decrease.
it can be seen that the slopes (both below and above the break energy) of the particle distribution increase with $p$. For the `non-contact' case ($\Rc>\Rs+\Delta\Rc/2$, e.g. $\Delta\Rc=8$~pc), the slopes above the break are very similar to those of the `contact' case, and the low energy cut-off becomes more prominent with the increase of $p$.
%, but the slope changes in  break becomes more gradual when $p$ decrease, with the break energy shifting to lower energy.

Figure~1{\em c} shows the spectral dependence on the diffusion coefficient index $\delta$ (with $\chi=0.01$, $\Delta R=15$pc, and $p=2.2$ fixed). The value of $\delta$ is usually adopted in the range of 0.3--0.7 (Berezinskii \etal\ (1990)). Here we use three values of $\delta$: $\delta=$0.3, very close to the Kolmogorov type diffusion index (0.33); $\delta=$0.5, for the Kraichnan type diffusion; and $\delta=$0.7. This parameter strongly influence the spectral pattern. The slope of the proton spectrum above the  break conforms the relation $\alpha_2\sim p+\delta$ for the continuous point source injection case (Aharonian 2004). For the `contact' case ($\Delta\Rc=10$~pc), there exhibits a ``platform" pattern extended to low energy, while for the `non-contact' case ($\Delta\Rc=8$~pc), the low energy cut-off becomes more significant with the increase of $\delta$.
%For $\chi=0.01$, $\Delta R=15$pc, and $p=2.2$, the top right panel of Fig.~\ref{fig:proton} shows different spectra for $\delta=$0.3(), 0.5, 0.7 (the value of $\delta$ is in the range of 0.3--0.7 (Strong \etal\ 2007) depending on different diffusion mechanism in different environments.).% The value of $\delta$ is in the range of 0.3--0.7 (Strong \etal\ 2007) depending on different diffusion mechanism in different environments. So three values of $\delta$ are used to investigate $\delta$-dependence with fixed $p=2.2$. This parameter indeed influence the width of the platform and the spectra slope after the break. %Empirical relation between the slope of the GeV platform ($\alpha_{\rm pf}$) and TeV spectrum ($\alpha_{\rm TeV}$) can be described as%\begin{equation}\label{eq:slope-pf} \alpha_{\rm pf}\sim p+0.25\delta-0.33?? \end{equation}

%Impacts of diffusion coefficient $\chi$
As shown in Figure~1{\em d}, different correction factor of diffusion coefficient $\chi$ gives different flux and width of the ``platform". With the increase of the $\chi$ value, the number of the low energy protons increases and the number of the high energy protons decreases, namely, the particle distribution gets softened. This is because faster diffusing high energy protons diffuse into farther distance and the low energy ones become dominant at the concerned distance. Meanwhile, the width of ``platform" becomes narrower and the slope of the proton spectra above the break energy keeps unchanged.

%It should be noted that the slope of proton spectrum before the break $\alpha_1$ is influenced by $p$, $\delta$ and the evolution of SNR, and the slope after the break $\alpha_2\sim p+\delta$, which is the same as the relation of the continuous point source diffusion case. (Aharonian book)

\begin{figure*}\label{fig:proton}
\begin{center}
\includegraphics[width=8cm]{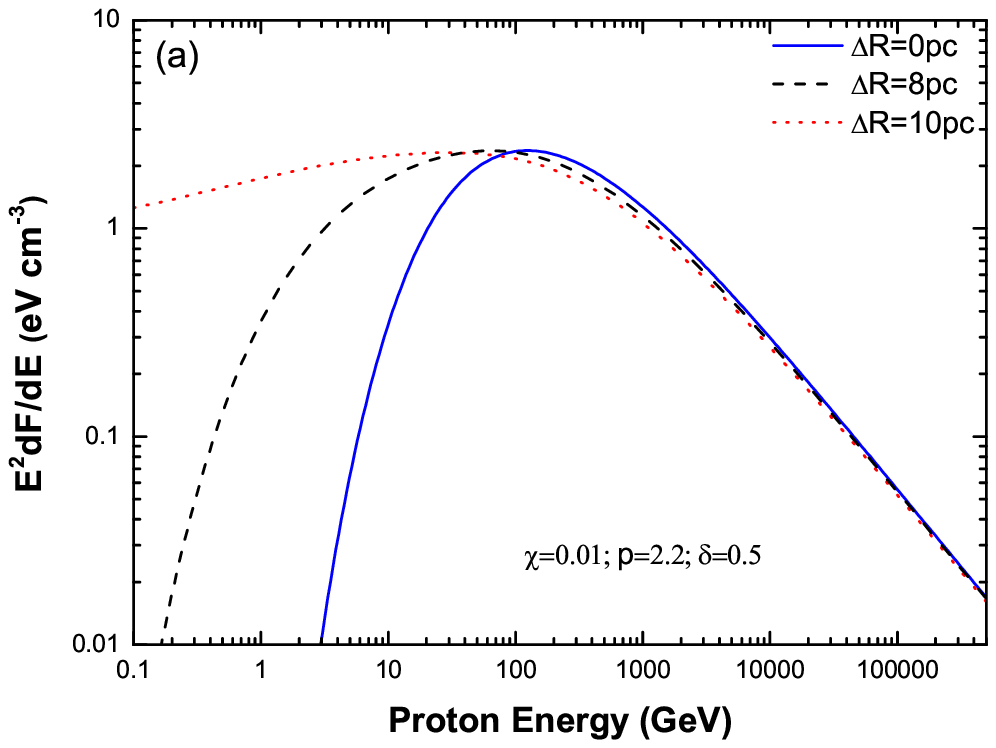}
\includegraphics[width=8cm]{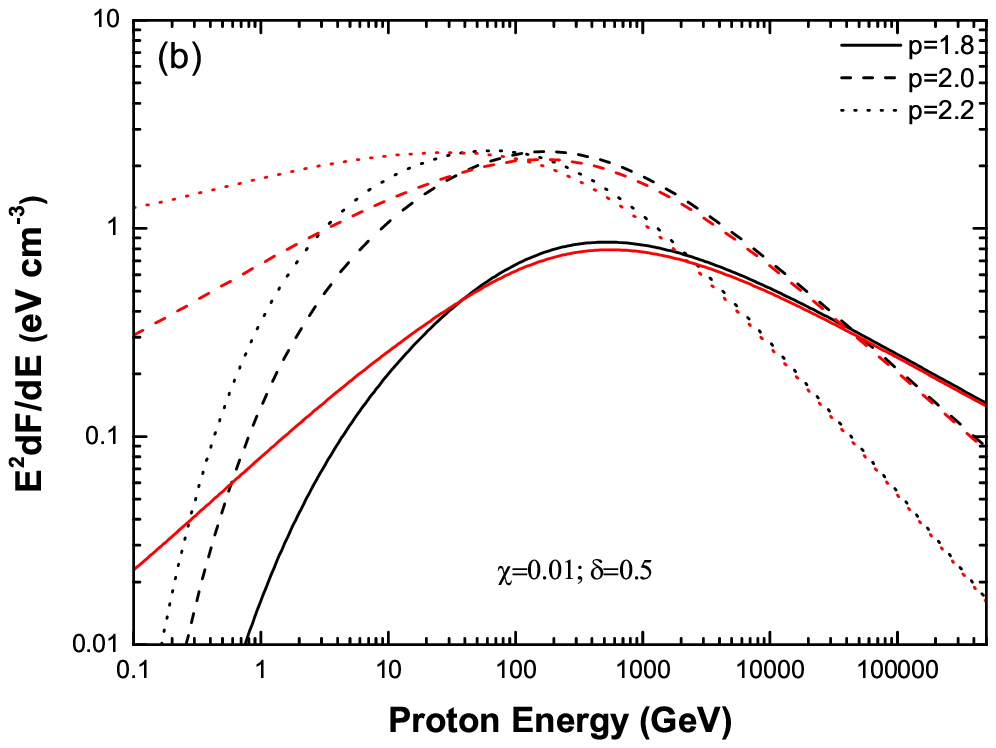}
\includegraphics[width=8cm]{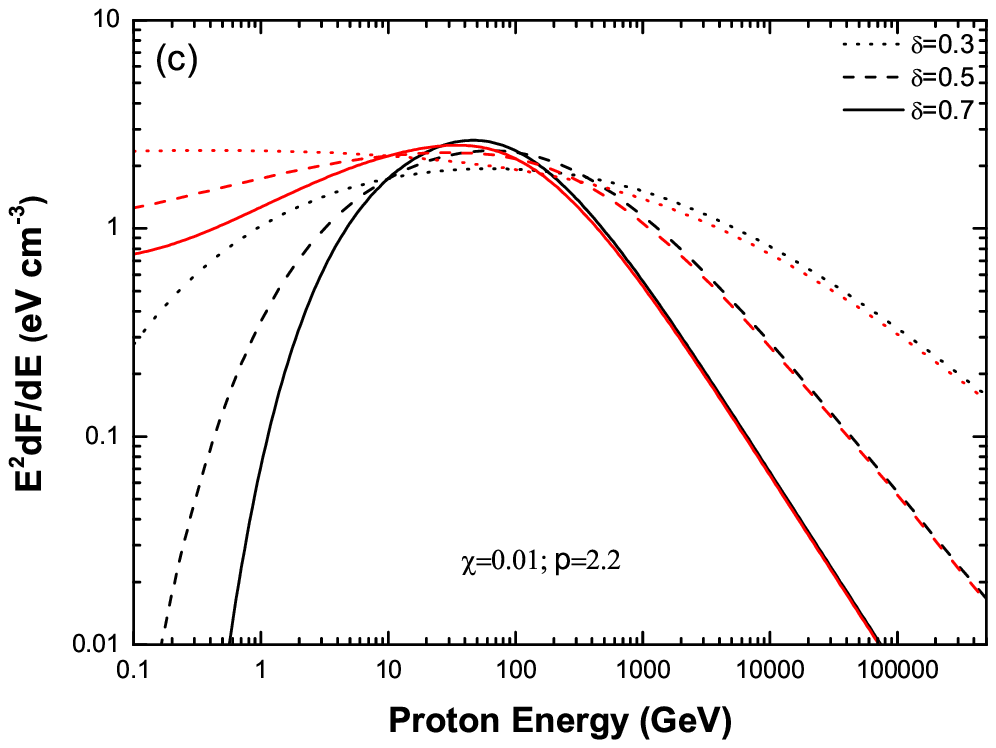}
\includegraphics[width=8cm]{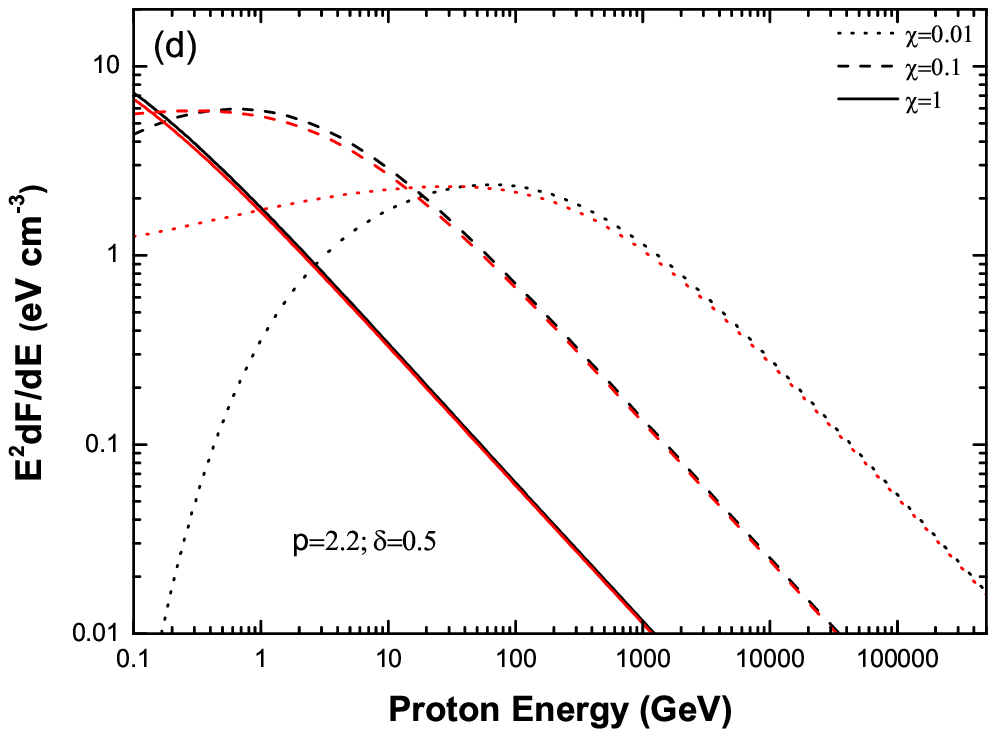}

\caption{Spectra of the diffusive protons in the MCs for $R_{\rm SNR}=20$pc, $t=27000\yr$, $\no=1\cm^{-3}$, $\ESNR=10^{51}$erg, and $\Rc=25$pc. In panel (a), $\Delta \Rc=0$pc (blue), 8pc (black), and 10pc (red) are adopted, respectively. In panels (b)-(d), the black curves represent the case for $\Delta \Rc=8$pc and the red curves for $\Delta \Rc=10$pc.}% In panel (b), $p=$1.8, 2, and 2.2 are adopted. In panel (c), $\delta=$0.3, 0.5, and 0.7 are adopted. Panel (d) shows the $\chi$ dependence of proton spectra with $\chi=$1, 0.1, and 0.01 are used.}
\end{center}
\end{figure*}

\section{Application to Individual SNRs}\label{sec:app}

Extended GeV emission associated with interacting SNRs has recently been revealed by a series of Fermi-LAT observations (Abdo \etal\ 2009, 2010a,b,c,d; Castro \& Slane 2010), some of which are also spatially resolved by the H.E.S.S. TeV observations. These SNRs form a special interesting class, which may be beneficial to unveil the enigma of the origin of the Galactic CRs.
One of the most conspicuous features of such a class is the commonly present, often gentle, spectral break at around 1-10GeV, which is difficult to be reproduced by leptons via inverse Compton emission (e.g. Abdo \etal\ 2010a). In addition, the TeV spectra are mostly steep, with the photon indices sometimes as high as $\sim3.0$, which is also hard to be explained by the primary accelerated protons, but can be done by the proton diffusion scenario (e.g. Aharonian \& Atoyan 1996; Torres \etal\ 2008; paper I; Ohira \etal\ 2011). Here we apply our improved accumulative diffusion model to the nine SNRs currently available of the class (W28, W41, W44, W49B, W51C, Cygnus Loop, IC443, CTB 37A, and G349.7+0.2) and reproduce the GeV-TeV spectral patterns as aforesaid.

The basic parameters of each SNR studied here are listed in Table~1 with $\ESNR=10^{51}$erg and $\eta=10\%$ assumed. Since they are all interacting SNRs, we only consider the MC which is in touch with the shock surface (i.e., the `contact' case, although in some cases not all the \grays\ is coming from a contact MC). Thus we adopt the value of the lower-limit of the integration of Eq.~(\ref{eq:finite-volume}) just the same as the value of the current shock radius, i.e., $\Rc-\Delta \Rc/2=\Rs$.

In the \gray\ spectral fit, the parameters are adjusted in the following way. Given a certain cloud mass (chiefly adopted from observation), the \gray\ flux can be roughly determined by changing the diffusion correction factor $\chi$ (e.g., for point-like injection case, $F_{\gamma}\propto\chi^{-3/2}$ (Gabici \etal\ 2010)). The broadness of the platform is fitted by adjusting the cloud thickness $\Delta\Rc$ for the fixed $\chi$ and the slope of the ``platform" and the slope above the break together with the TeV data can in turn be obtained with proper values of $p$ and $\delta$.

%Before fitting the \gray\ spectra, we carefully examine the literatures of these SNRs to determine their distances, current radius, ages in order to establish their dynamical evolution. These parameters are listed in Tab.~\ref{tb:dyn}. The most important and uncertain parameter is the distance of SNRs.

%Hewitt \etal\ (2009) show that the OH masers which strongly trace the shock-MC interaction have a strongly correlation with \gray\ sources which strongly support the scenario that these \grays\ are hadronic origin. What's more, after survey these potentially interacting GeV SNRs, we find that, the SNRs all have the universal spectra break or platform before the break.

%Do survey on the Fermi sources are find the SNR-MC interacting system, we find it interesting that the GeV break appears universally. In addition, For the uncertainty of distance and age of these sources, the rough estimation should be made and the by dynamical evolution parameters we used are listed in Tab.~\ref{tb:dyn}.

\begin{table*}\label{table:dyn}
\begin{center}
\caption{Dynamical evolution parameters for interacting SNRs}
\begin{tabular}{cccccccc}
\hline
SNR                 & age(kyr)      & distance(kpc) & angular size & $R_{\rm SNR}$(pc) & Reference \\
\hline
G6.4-0.1(W28)       & 42            & 1.9           & $48'$        & 13.3              & 1, 2, 3\\
G23.3-0.3(W41)      & 100           & 4.2           & $33'$        & 20.2              & 4 \\
G34.7-0.4(W44)      & 20            & 2.8           & $30'$        & 12.2              & 5,6 \\
G43.3-0.2(W49B)     & 2             & 10            & $3'$         & 4.4               & 7\\
G49.2-0.7(W51C)     & 30            & 6             & $30'$        & 26.2              & 8,9\\
G74.0-8.5(Cygnus Loop)&20           & 0.54          & $180'$       & 14.1              & 10,11,12\\
G189.1+3.0(IC443)   & 20            & 1.5           & $45'$        & 9.8               & 6\\
G348.5+0.1(CTB 37A) & 30            & 11.3          & $15'$        & 24.7              & 13,14\\
G349.7+0.2          & 2.8           & 23            & $2'$         & 6.7               & 14,15\\
%Cas A & 0.33 & 3.4 & 5' & 2.5 & 0.15 & Sedov \\
\hline
\end{tabular}
\end{center}
\medskip
References.--(1) Paper I; (2) Vel$\acute{\rm a}$zquez \etal\ 2002; (3) Green 2009; (4) Tian \etal\ 2007; (5) Abdo \etal\ 2010a; (6) Seta \etal\ 1998 (7) Brogan \& Troland 2001; (8) Koo \etal\ 2005; (9) Moon \& Koo 1997a; (10) Green 1990; (11) Blair \etal\ 2005; (12) Miyata \etal\ 1994; (13) Sezer \etal\ 2011; (14) Reynoso \& Mangum 2000; (15) Slane \etal\ 2002
\end{table*}

\begin{table*}\label{table:para}
\begin{center}
\caption{Fitted parameters for the \grays\ of interacting SNRs}
\begin{tabular}{cccccccc}
\hline
SNR                & $p$ & $\delta$ & $\chi$ & $\Delta$ R(pc) & $M_{cl}$ ($10^4M_{\sun}$)\\
\hline
G6.4-0.1(W28)      & 2.2 & 0.5      & 0.05   & 0.5            & $5^a$                   \\
G23.3-0.3(W41)     & 2.1 & 0.6      & 0.004  & 17             & $7^b$                 \\
G34.7-0.4(W44)     & 2.4 & 0.7      & 0.02   & 7              & $3^c$                 \\
G43.3-0.2(W49B)    & 2.4 & 0.6      & 0.03   & 3              & $1^d$               \\
G49.2-0.7(W51C)    & 2.2 & 0.5      & 0.04   & 7              & $1^e$                  \\
G74.0-8.5(Cygnus Loop)&2.2& 0.7     & 0.04   & 8              & $0.025^f$               \\
G189.1+3.0(IC443)  & 2.3 & 0.65     & 0.01   & 7              & $0.5^g$                  \\
G348.5+0.1(CTB 37A)& 2.1 & 0.6      & 0.02   & 5              & $5.8^h$                 \\
G349.7+0.2         & 2.1 & 0.7      & 0.02   & 3              & $1.2^i$                  \\
%Cas A & 0.33 & 3.4 & 5' & 2.5 & 0.15 & Sedov \\
\hline
\end{tabular}
\end{center}
\medskip
References.--(a) Aharonian \etal\ 2008a (b) Seta \etal\ 1998; (c) Seta \etal\ 2004; (d) Assumed; (e) Koo \& Moon 1997 a,b; (f) Assumed (g) Cornett \etal\ 1977; (h) Reynoso \& Mangum 2000; (i) Reynoso \& Mangum 2001

\end{table*}

\subsection{G6.4-0.1 (W28)} \label{W28}

W28 is one of the prototype thermal composite (or mixed-morphology) SNRs, characterized by shell-like radio emission and centre-brightened thermal X-rays. It is an evolved SNR in the radiative phase, with an age estimate ranging from 35 to 150 kyr (Kaspi \etal\ 1993). %The observations of molecular lines place it at a distance of $\sim2$kpc (Vel¡äazquez et al.2002).
In \grays\ the strongest TeV source is located at the northeastern edge of the remnant and positionally coincident with the well studied MC
with which the remnant shock interacts (Aharonian \etal\ 2008a). In this \gray\ emitting region, there are some of the OH masers points detected in W28 (Frail \etal\ 1994). Recently, \Fermi\ LAT observation found that the GeV source 1FGL J1801.3-2322c coincides with the TeV source and thus demonstrates a broken power-law for the \gray\ spectrum with a break at $\sim1$ GeV and photon indices of $\sim2.09$ below the break and $\sim2.74$ above the break (Abdo \etal\ 2010c). Three molecular clumps (Aharonian \etal\ 2008a) located outside the southern boundary of the remnant have different GeV and TeV brightness. In Paper I, we have explained the \gray\ spectra of the four sources via hadronic process of the diffusive protons, assuming different distances for the four sources. Here we also fit the spectrum of the northeastern source, taking the cloud volume into account. The result is very similar to that in Paper I. Unlike the sources in other SNRs studied in this paper, the ``platform" below the break is unapparent (Figure~\ref{fig:cases}(a)), entailing a large value of $\chi$ and a small cloud thickness $\Delta \Rc$. The evaluation of $\chi$ also sensitively depends on the cloud mass.

\subsection{W41 (G23.3-0.3)}\label{W41}
HESS J1834-087 used to be representative of a class of ``dark" \gray\ sources, which was not thought to be firmly coincident with any detected sources (Aharonian \etal\ 2006). New H.E.S.S. analysis has revealed that the TeV emission can be divided into two components: the compact and extended ones. While the TeV compact component compatible with pulsar candidate does not seem to have GeV counterparts, the extended one, with index 2.7, may spatially match the GeV emission detected by \Fermi-LAT from the SNR W41 region (M\'ehault \etal\ 2011), making this SNR a promising candidate for generating the observed extended \grays. The \gray\ spectrum in \Fermi\ regime can be fitted by a single power-law with index 2.1, which causes a break between GeV and TeV.

This field is also observed in radio with VLA, CO, and X-rays with XMM-Newton. The high-resolution CO observation reveals its association with a giant molecular cloud with mass $>10^4M_{\sun}$ (Tian \etal\ 2007). The strong absorption in X-rays together with the high the \gray\ to X-ray luminosity ratio ($\sim11$) indicates that the \grays\ may be of hadronic origin from energetic protons colliding with shocked GMC (Tian \etal\ 2007; Yamazaki \etal\ 2006).

The GeV ``platform" of this source is the broadest ($\sim 1-100$GeV) among the SNRs studied here (see Figure~\ref{fig:cases}(b)). It thus requires a low diffusion coefficient ($\chi\la 10^{-2}$) and large thickness of cloud ($\Delta \Rc$ comparable to $\Rs$). With the parameters used (see Table~2), the well-determined TeV slope is also reproduced.

\subsection{W44 (G34.7-0.4)} \label{W44}
W44 is also a paradigm of ``mixed-morphology" SNR, with semi-symmetric shape in radio with $\sim30'$ in size (Rho \& Petre(1998)).
%[Based on HI absorption observation, the distance of W44 is estimated to be 2.8kpc which correspond a radius of 12pc (...).]
It is considered to interact with a dense molecular cloud, with evidences including OH maser (Hoffman \etal\ 2005), molecular line broadening, CO line ratio (Castelletti \etal\ 2007), etc.
%It is shown the coincidence of the brightest radio filaments obtained from 324 MHz observation with the 4.5 $\mu$m infrared emission with the exception of central and eastern regions . OH (1720 MHz) maser emission found in this source also supports the physical interaction of the blast wave with the clumpy ISM .

The \gray\ spectral energy distribution of W44 represents a broken power-law, which breaks at 1.9GeV, with photon indices 2.06 at low energy and 3.02 at high energy. Most of the \gray\ emission is inferred to arise from the SNR other than from the unresolved pulsar based on the resemblance between the \gray\ and infrared ring morphologies (Abdo \etal\ 2010a). Most recently, AGILE observation has also been reported with the \gray\ spectral down to 50 MeV. (Giuliani \etal\ 2011)

Figure~\ref{fig:cases}(c) shows the fitting results of W44 using parameters listed in Table~2. Our fitting not only well reproduce the GeV ``platform" spanned from 0.3 to 1.9 GeV, but also the spectrum with energy as low as 50 MeV, which is a robust advantage for hadronic origin of the \grays. It is noteworthy that the photon index above the break energy (3.02) is the largest among the 8 studied SNRs; we adopt the upper limit (0.7) of the usual range for $\delta$ so as for $p$(=2.4) not to deviate too much from the common values (1.8-2.2). Although the VHE spectrum of W44 is the very steep, we predict that the flux around TeV energy is slightly higher than the H.E.S.S. detection limit and much higher than the expected sensitivity of new generation TeV telescope Cherenkov Telescope Array (CTA) (de Cea del Pozo \etal\ 2009).

\subsection{W49B (G43.3-0.2)} \label{W49B}
W49B is a ``mixed-morphology" SNR at a young age. %The distance is estimated to be 8-11kpc (Radhakrishnan \etal\ 1972; Moffett \& Reynolds 1994; Brogan \& Troland 2001) which corresponds to a SNR Radius 5-7pc.
Near-infrared shocked H$_2$ emission suggests that W49B is an interacting SNR (Keohane \etal\ 2007). Particularly high abundances of hot Fe and Ni, and relatively metal-rich core and jet regions are interpreted as evidence that W49B is originated inside a wind-blown bubble inside a dense molecular cloud (Keohane \etal\ 2007).

Recently, \Fermi\ LAT reported a GeV source spatially coincident with W49B (Abdo \etal\ 2010d). The spectrum, again, exhibits a broken power-law shape with break energy $\sim4.8$GeV and the simple power-law can be rejected at a significance of 4.4$\sigma$. More interestingly, the best-fit spatial position of the LAT source which seems coincident with the brightest part of the \Spitzer\ IRAC 5.8$\mu$m image strongly supports the scenario that the GeV emission is linked with the shocked gas. % On the other hand, \gray\ emission had not been detected in the TeV bands.
Recently, the H.E.S.S. collaboration has reported its detection in TeV and the photon spectrum is smoothly connected with GeV data points (Brun \etal\ 2010).

Due to the lack of definite value of the molecular mass around this SNR, we assume $M_{\rm cl}\sim10^4M_{\sun}$ for fitting the \gray\ spectrum. The spectral slope below the break energy ($<4.8$GeV) is relatively large ($\sim2.2$) and, consequently, a large value of $p$ ($=2.4$) is inferred.

\subsection{W51C (G49.2-0.7)} \label{W51C}
W51C is also a member of ``mixed-morphology" type of SNRs. A partial shell $\sim30'$ in diameter with a breakout feature in northern part appears in the radio continuum (Moon \& Koo 1994).
%The distance of this SNR is approximated as 6kpc by CO observation (??) which indicate a radius $\sim30pc??$. The age ($\sim3\times10^4$yr) and explosion energy ($\sim3.6\times10^51$erg) were estimated for either the Sedov or evaporative models (Koo 1995).
A shock-MC interaction was shown by the observations of the shocked atomic and molecular gases (Koo \& Moon 1997 a,b).

In the \grays, the H.E.S.S. collaboration has detected an extended VHE source coincident with W51C (Fiasson \etal\ 2009), and the Milagro observation reported a possible excess in this direction (Abdo \etal\ 2009). Recently, \Fermi\ LAT detects the flux in this field and shows that the GeV spectrum has a ``platform" below the break energy $\sim2$GeV and cannot be described by a single power-law, which makes it difficult for leptonic models to reproduce the GeV spectrum (Abdo \etal\ 2009).
Fang \& Zhang (2010) explain the GeV-TeV spectrum with hadronic process, using the non-linear acceleration by low Mach number shock, and in this scenario the \grays\ arise from the acceleration region.

Considering the protons diffusion into the adjacent MC, we perfectly reproduce the GeV-TeV spectrum with typical parameters (e.g., $p=2.2, \delta=0.5$) as shown in Figure~\ref{fig:cases}(e).

\subsection{Cygnus Loop (G74.0-8.5)} \label{Cygnus Loop}
%The Cygnus Loop is one of the most famous and well-studied middle-aged SNRs. The X-ray emission from thermal plasma is prominent in the shell of the remnant, with a blowout from the southern rim (Ku et al. 1984). The blast wave on the western limb might encounter molecular material (Scoville et al. 1977), while on the northeast part ... (Graham \etal\ 1991). The Cygnus Loop is interacting with a protrusion of the cavity wall in its eastern edge (the XA region), where the X-ray emission is very bright.(Zhou \etal\ 2011). Most recently, \Fermi\ LAT observation have detected significant \gray\ emission associated with this SNR in the energy band 0.2--100 GeV and the spectrum shows a break in the range 2--3 GeV. The gamma-ray luminosity is $\sim1\time10^{33}erg~s^{-1}$, which is much lower than those of other GeV-emitting SNRs. The spatial distribution of the \grays\ is best represented by a ring shape, with inner/outer radii 0.7 and 1.6. (Katagiri \etal\ 2011).

The Cygnus Loop is one of the most famous and well-studied middle-aged SNRs. The \gray\ emission mapped by Fermi-LAT is along the SNR limb and appears to be broken into four parts, three of which are essentially in correspondence with H$\alpha$ bright patches (Katagiri \etal\ 2011).  Actually, there have been some evidence of the presence of dense gas along the SNR boundary. In the west, CO emission in $-25$-- +30 km~s$^{-1}$ interval seems to be adjacent to the SNR (Dame 2001, see Katagiri \etal\ 2011) and two CO emission clumps at $\sim+9.2$ km~s$^{-1}$ (with a mass $\sim100M_{\sun}$) seems to be well in contact with the optical filament (Scoville et al.\ 1977). Shock excited near-IR $\rm H_{2}$ line emission associated with the SNR is detected from the northeastern boundary (Graham \etal\ 1991a,b). Also, according to the X-ray studies, dense wall of cavity is suggested to be along the eastern edge (Levenson \etal\ 1997;
Levenson, Graham \& Snowden 1999) and shock-cloud interaction is thought to be taking place there (e.g., Miyata \& Tsunemi 2001; Zhou \etal\ 2010).  Such a molecular-rich environment may act as a reasonable site for the hadronic origin of \gray\ emission from shock accelerated CRs, which is pointed out by Katagiri \etal\ (2011). %which is noticed from the \gray\ spectral fit by Katagiri \etal\ (2011).

\Fermi\ LAT observation in the energy band 0.2--100 GeV shows a spectral break in the range 2--3 GeV. The \gray\ luminosity is $\sim1\times 10^{33}erg~s^{-1}$, which is lower than those of other GeV-emitting SNRs. The \gray\ spectrum can be reasonably fitted by our model.%The spatial distribution of the \grays\ is best represented by a ring shape, with inner/outer radii 0.7 and 1.6(Katagiri \etal\ 2011).
The fitting result is shown in Figure~\ref{fig:cases}(f) with parameters listed in Table~2. To specify $\Delta\Rc$, we adopt the inner and outer radii $0.7'$ and $1.6'$ of the \gray\ ring-like structure (Katagiri \etal\ 2011). %A small mass () is required due to the low \gray\ luminosity.% which is similar to the CO observations.

\subsection{IC443 (G189.1+3.0)} \label{IC443}
IC 443 is one of the most thoroughly studied interacting SNRs, with rich evidence including the OH masers from the central and southeast regions (Hewitt \etal\ 2006), CO line ratio (Seta \etal\ 1998), molecular line broadening (Dickman \etal\ 1992), near-infrared shocked H$_2$ emission (Rosado \etal\ 2007), etc.
%The SNR has an angular extent of $\sim45'$ in the radio with a complex shape consisting of two half-shells with different radius.
%The HII region S249 interacting with the northeast shell of IC443 place it at a distance of 1.5kpc (Fesen 1984) and the corresponding radius of this SNRs is roughly 9.5pc in average. The age of IC 443, however, remains uncertain in the range of $\sim3-30$kyr (VERITAS and references there in).

In \grays, there are several sources coincident with IC 443 as observed by EGRET (Hartman \etal\ 1999), AGILE (Tavani \etal\ 2010), and \Fermi\ (Abdo \etal\ 2010b) in GeV and MAGIC (Albert \etal\ 2007) and VERITAS (Acciari \etal\ 2009) in TeV. The spectrum can be represented by a broken power law with slopes of 1.93 and 2.56 and with a break at 3.25GeV and show very broad ``platform" below the break (Figure~\ref{fig:cases}(f)).

The multi-band photon spectra, with only the EGRET data then available in GeV band, was explained by using a time-dependent model for non-thermal particles emitting in the acceleration region (Zhang \& Fang 2008). Torres \etal\ used the point source diffusion model assuming two clouds (with volumes ignored) in the vicinity of IC443 to explain the GeV-TeV connection (Torres \etal\ 2010). %Ohira \etal\ developed a diffusion model, assuming Sedov phase SNR, and explained the spectral shape of this SNR without considering the flux.

As shown in Figure~\ref{fig:cases}(f), the \gray\ spectrum of IC443 is fitted assuming one cloud with geometric volume.  The broad GeV ``platform" is reproduced with a low diffusion coefficient ($\chi\sim10^{-2}$) and a large thickness of cloud required.

\subsection{CTB 37A (G348.5+0.1)} \label{CTB37A}
SNR CTB~37A has a partial shell with a extended breakout to the south (Kassim \etal\ 1991). %The distance of this SNR is approximated as 10.3kpc by using HI absorption (Caswell \etal\ 1975) , while the CO observation place it at a distance of 11.3kpc which implies a size of the remnant close to 28pc(Reynoso \& Mangum 2000).
The 1720~MHz OH masers (Frail \etal\ 1996) and morphologically correspondent CO emission (Reynoso \& Mangum 2000) demonstrate that this SNR is interacting with MCs.

In high energy \grays\ , H.E.S.S. has detected TeV emission in the field of CTB 37A. For the absence of X-ray synchrotron emission, it is unlikely that the leptonic scenario could explain these VHE emission and a hadronic origin would be favorable (Aharonian \etal\ 2008b). Recently, an unresolved source lies within the eastern shell of CTB 37A was detected by \Fermi\ LAT, and the spectrum in \Fermi\ regime, too, shows a ``platform" below $4.2$~GeV (Castro \& Slane 2010). The GeV-TeV spectrum is well fitted (see Figure~\ref{fig:cases}(g)) with parameters listed in Table~2.

\subsection{G349.7+0.2} \label{G349}
G349.7+0.2 is one of the radio and X-ray brightest SNRs in the Galaxy, with an blowout morphology (Slane \etal\ 2002). It has been shown to be an interacting SNR, with numerous evidences including the OH masers (Frail \etal\ 1996), CO line ratio of different transitions, molecular line broadening (Dubner \etal\ 2004), near-infrared(IR) shocked H$_2$ emission (Reach \etal\ 2006), etc.

\begin{figure*} %[ht]
\begin{center}
\includegraphics[width=0.32\textwidth]{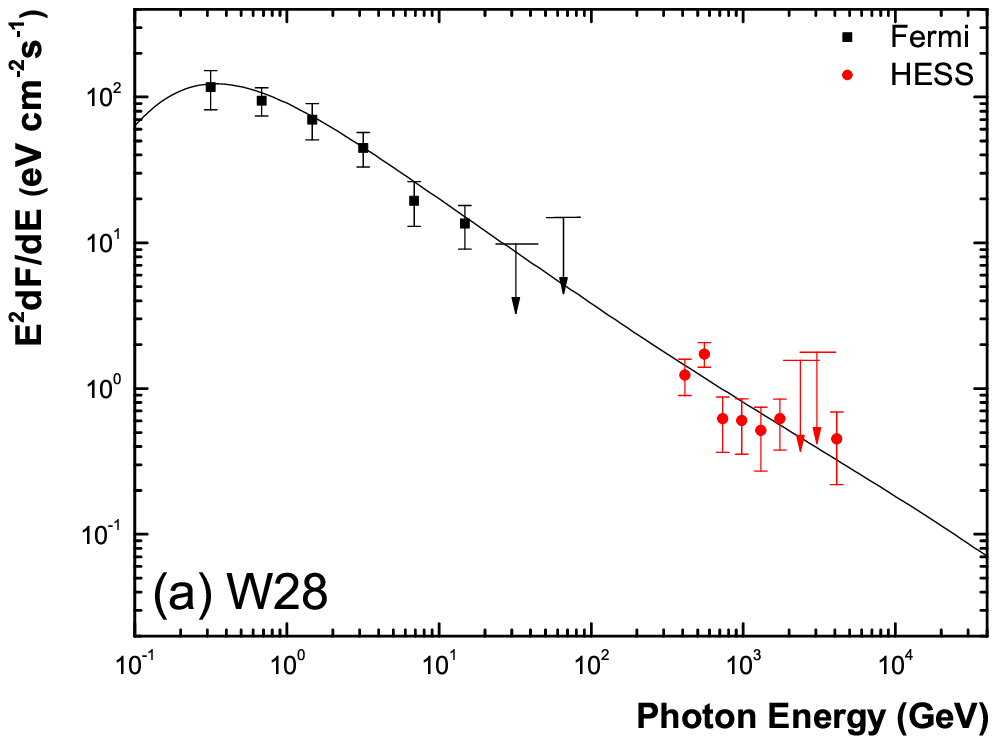}
\includegraphics[width=0.32\textwidth]{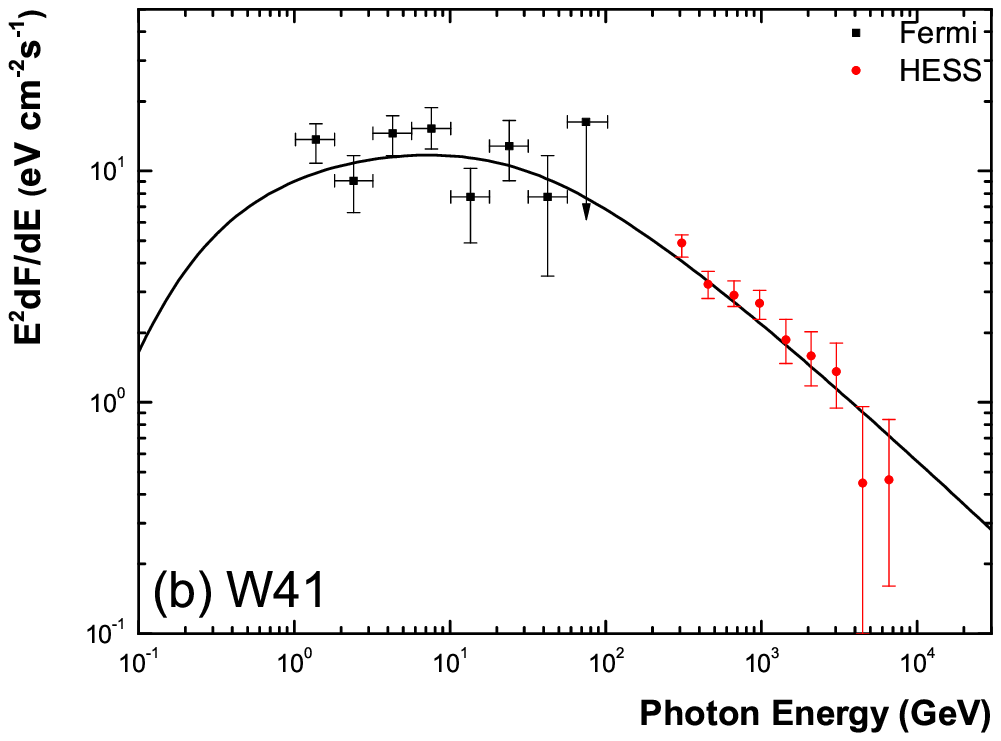}
\includegraphics[width=0.32\textwidth]{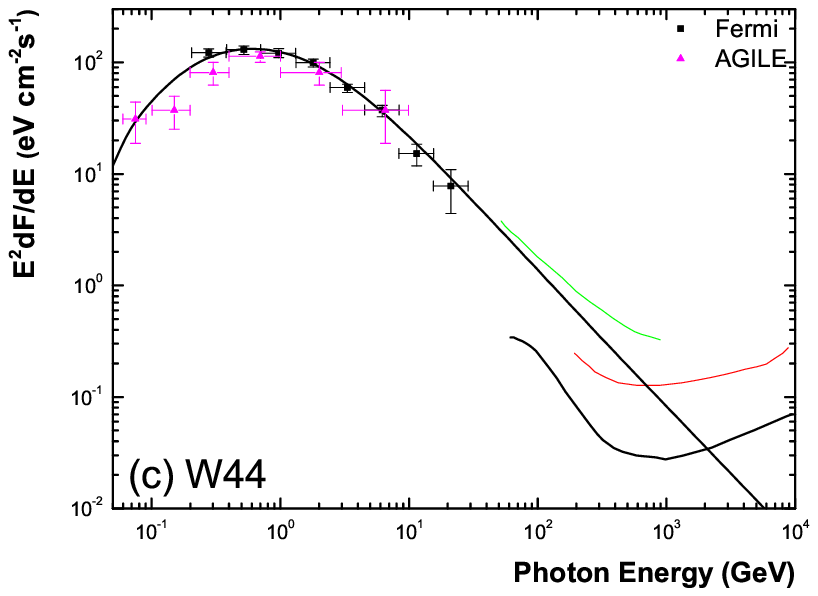}
\includegraphics[width=0.32\textwidth]{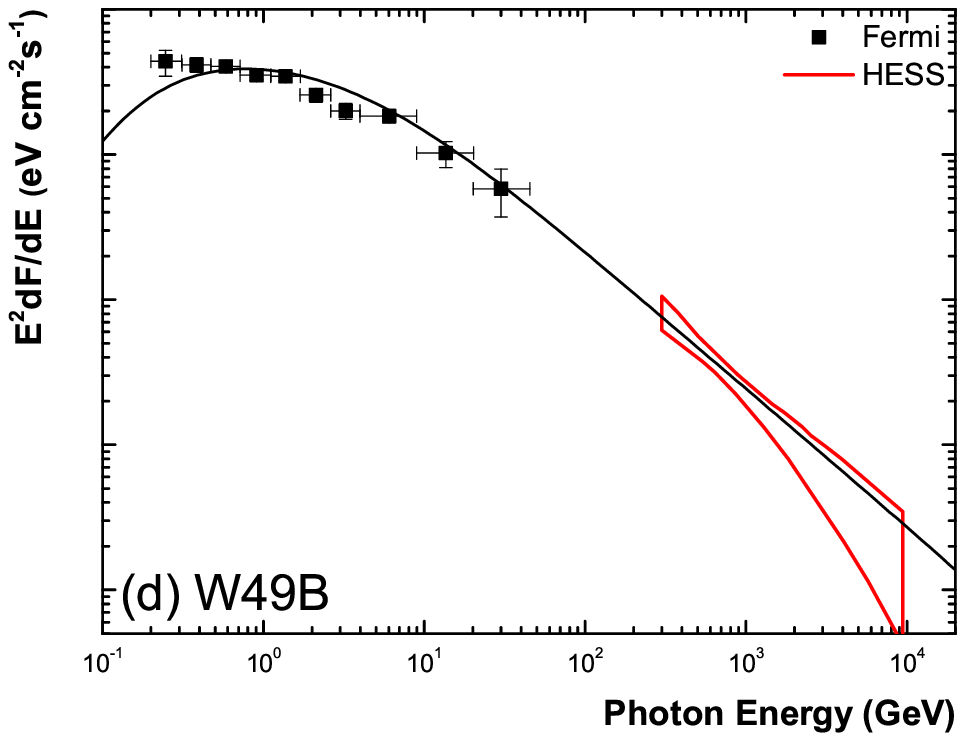}
\includegraphics[width=0.32\textwidth]{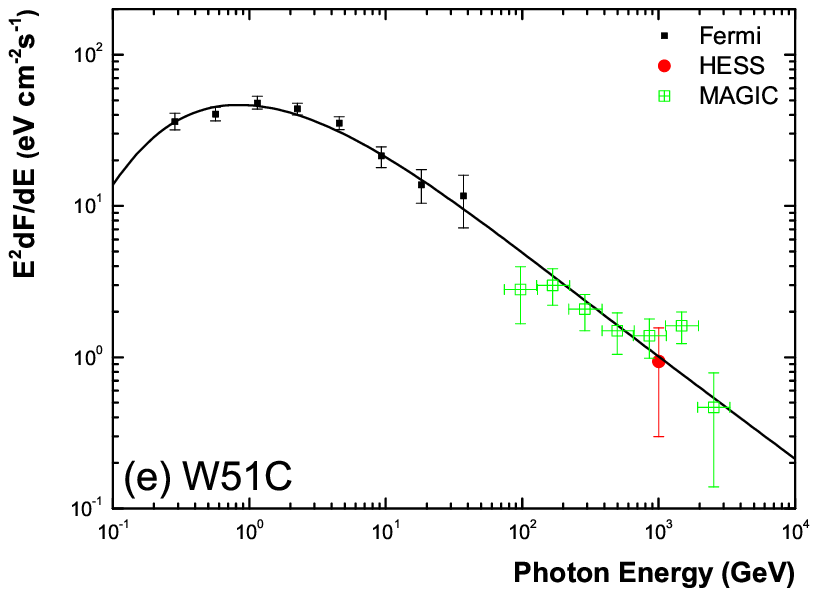}
\includegraphics[width=0.32\textwidth]{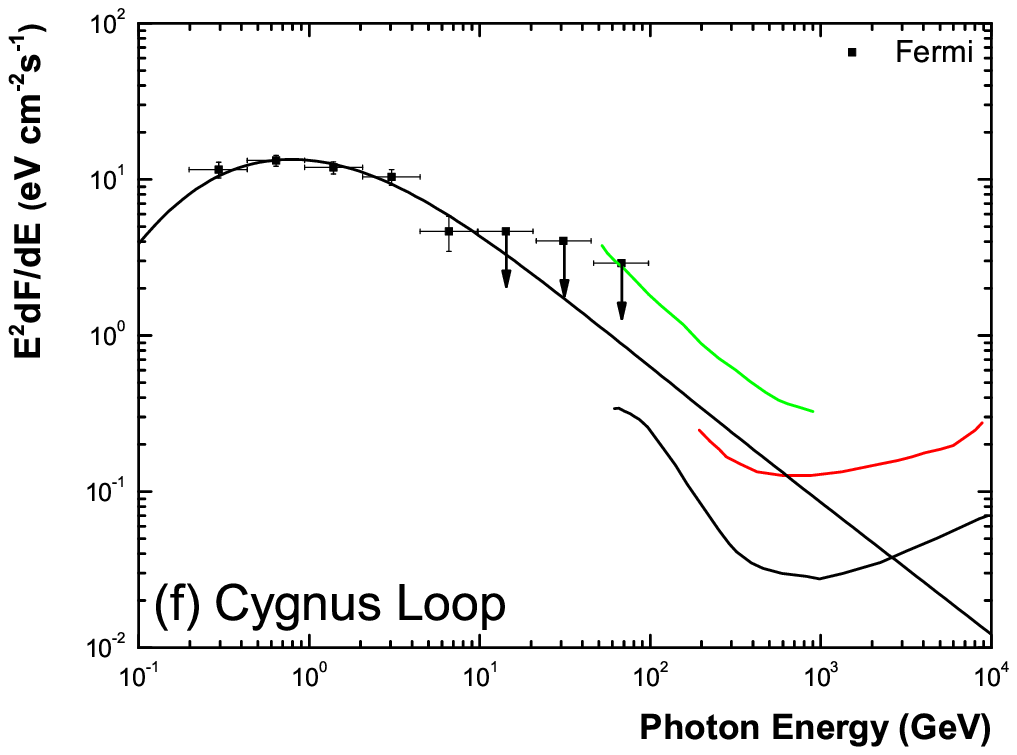}
\includegraphics[width=0.32\textwidth]{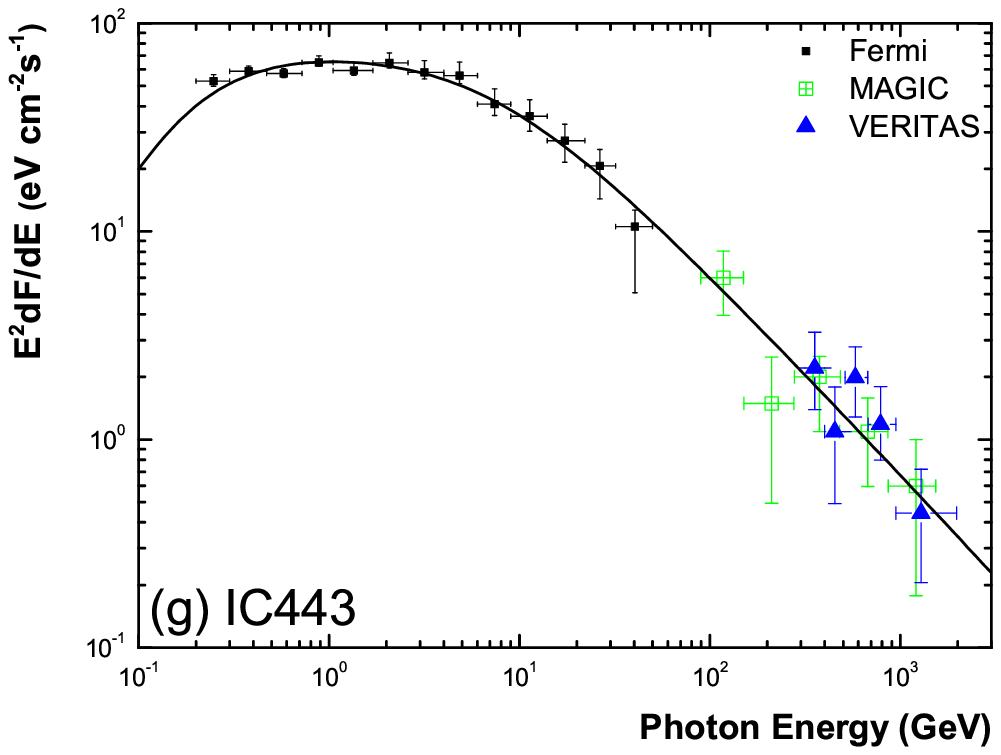}
\includegraphics[width=0.32\textwidth]{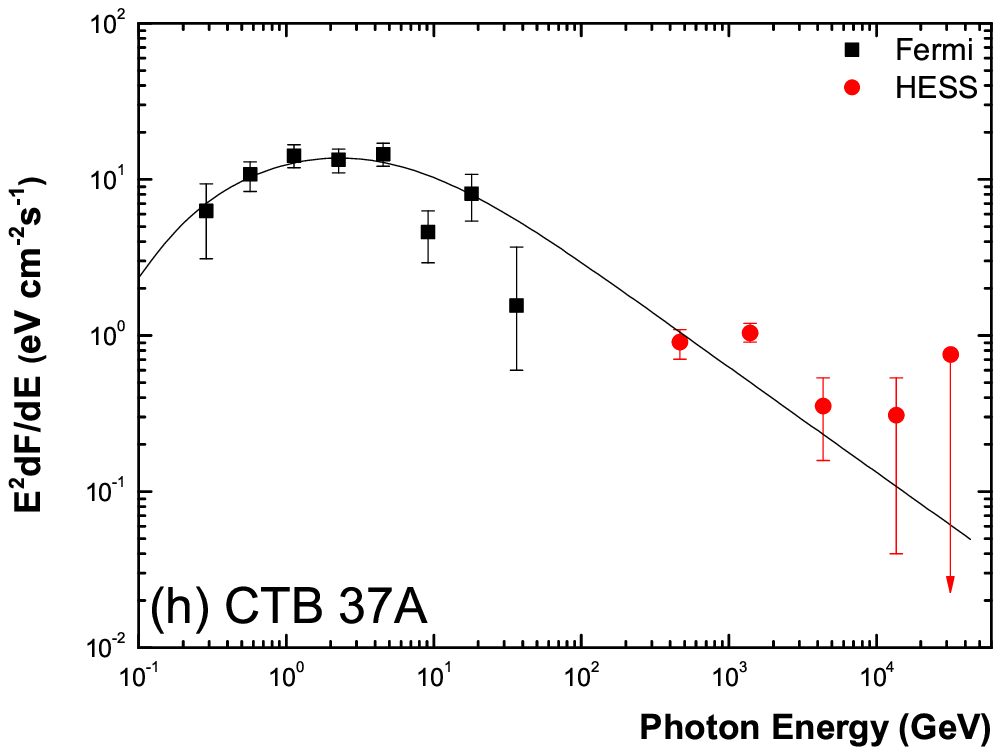}
\includegraphics[width=0.32\textwidth]{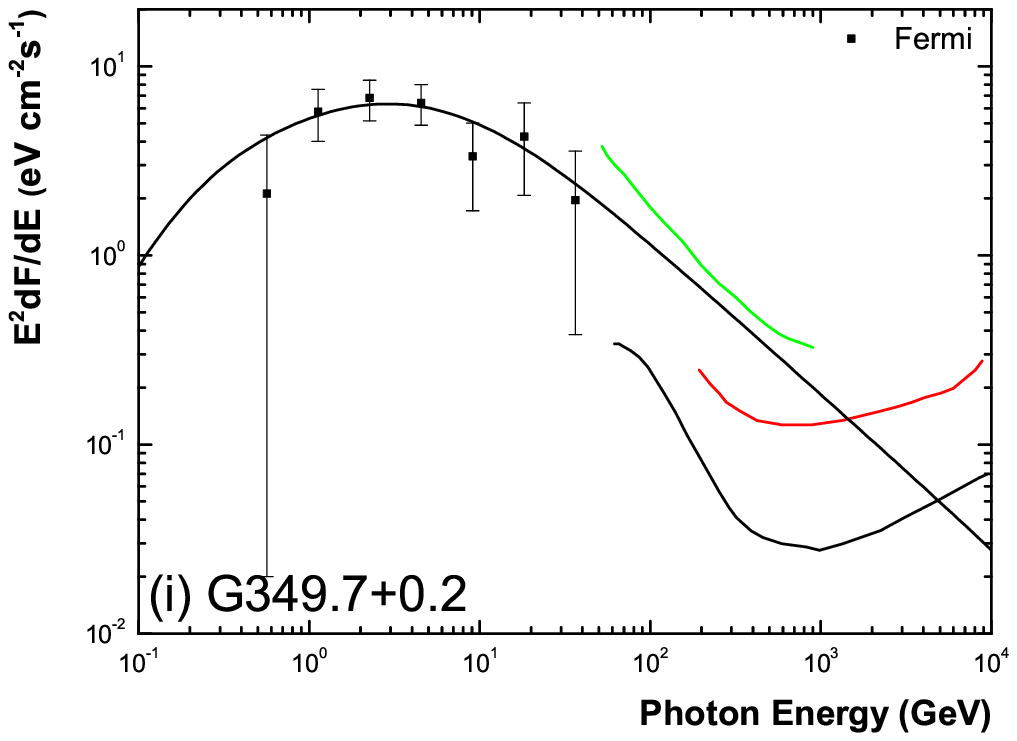}
\caption{\gray\ spectral data of the eight interacting SNRs and the model spectra. The black square and red dot show the data points observed by \Fermi\ and H.E.S.S., respectively. The sensitivity curves for MAGIC (green), H.E.S.S. (red), and CTA (black) are also shown in (c), (f) and (i). Panels from (a) to (h) represent the spectra fit of SNRs W28, W41, W44, W49B, W51C, Cygnus Loop, IC443, CTB 37A, and G349.7+0.2, respectively. (See text in Sec.~\ref{sec:app})}
\label{fig:cases}
\end{center}
\end{figure*}

The \Fermi\ LAT observation shows a bright \gray\ source in its field with a significance from the evaluation of the test statistic $\sim10\sigma$. The spectrum exhibits a broken power law with break energy $\sim16$GeV (Castro \& Slane). As yet, no TeV excess has been reported in this direction.
With the spectral fitting (see Figure~\ref{fig:cases}(h)), we predict that the flux around TeV energy is slightly higher than the H.E.S.S. detection limit and much higher than the expected sensitivity of CTA.

%\subsection{Cas A (G49.2-0.7)} \label{CasA}
%Cas A is one of the youngest known SNRs in our Galaxy.
%Hard to explain by pure leptons (Araya \& Cui). The GeV excess can be well %explained by additional \grays\ from hadronic.
%\subsection{3C391 (G49.2-0.7)} \label{3C391}

\section{Discussion and Summary}

In this paper, the accumulative diffusion model for CRs escaping from SNR shock (Paper I) is refined by considering the finite-volume clouds in the vicinity of SNRs. The variation of the proton spectra with different model parameters are also shown for exploring the parameter space. This refined model is applied to nine Galactic SNRs which are thought to be interacting with the ambient dense material, and their GeV spectral breaks/``platform"s, together with available TeV data, are naturally fitted.
%We have also found that the broadness of the spectral ``platform" is strongly influenced by the thickness of the clouds, escaping proton spectra and diffusion coefficient.

For the four SNRs (W51C, W28, W44 and IC443), Ohira \etal\ (2011) obtain the momentum breaks of proton spectra and successfully explain the \gray\ spectral shapes. In our model, in addition to fitting the spectral shapes, the real differential fluxes in \grays\ are also calculated, with the cloud masses considered.
%the cloud mass, we can fit not only the spectral shape of these interacting SNRs but also the absolute magnitude of the flux using the observed mass of the molecular cloud (except for the case of W49B for which the MC mass is hardly determined in observations).
For all nine SNRs, the parameter $\delta$, which gives the energy dependence of the diffusion coefficient, is used in an interval from 0.5 to 0.7, which is in the commonly favored range 0.3-0.7. The difference of the spectral fit parameters between our model and Ohira \etal's (2011) ($\delta=0.19-0.62$) may be primarily caused by two factors. The first factor is the different assumption of escaping spectrum of protons: in their model, the time evolution of maximum energy and the delta-function momentum of the escaping protons are assumed, and in our model, typical power-law distribution for escaping protons is used (e.g., Aharonian \& Atoyan 1996; Torres \etal\ 2008). The second factor is the relative positions between the shock and the \gray\ emitting MC: in their calculation, the MC is placed on the wall of stellar wind bubble separated from the shock (namely, the `non-contact' case), while in our calculation, the `contact' case is considered.

For fitting the GeV-TeV spectra of these interacting SNRs, the diffusion coefficients (see Table~2) are required to be much smaller ($\chi\sim 10^{-2}$) than the averaged Galactic value $\sim1$ and are also different from the values used in Ohira \etal\ (2011), which are in a wide range from 0.01 to 1. Such small values of $\chi$ have been noticed in observation by several authors (e.g., Fujita \etal\ 2009; Giuliani \etal\ 2010; Paper I; Gabici \etal\ 2010; Torres \etal\ 2011) when applying the proton diffusion to the \gray\ sources in the vicinity of SNRs. Theoretically, this phenomenon is also studied by Monte-Carlo simulation,
 %considering that the CR streaming generates Alfven waves around the SNR
%. According to the simulation,
in which the CR streaming instability can generate Alfv\'en waves. The waves strongly scatter the particles and make the particle diffusion in the ISM around the SNR remarkably slow, and the diffusion coefficient is therefore strongly suppressed (Fujita \etal\ 2010; Fujita \etal\ 2011). This kind of suppression is significant if the ambient ISM is well ionized for the cosmic rays around the SNRs, because the neutral damping of Alfv\'en waves is not effective.

For small $\chi$, the typical diffusion velocity for certain particle energy may be smaller than the shock velocity, %the "escape spectrum" cannot really be a power law, simply because
and the low energy cosmic rays will be continuously overrun by the SNR and reaccelerated. When considering particle acceleration, one needs a small Bohm diffusion coefficient ($D_{\rm bohm}$) to ensure particles to be proximate to the shock and repeatly cross it efficiently. The reason why particles escape is that their diffusion coefficient in the ISM ($D_{\rm esc}$) is much larger than the one near the shock, namely $D_{\rm esc}\gg D_{\rm Bohm}$. In our paper, even if the diffusion coefficient is about two orders of magnitude smaller than the Galactic average, this condition is satisfied well. Even if some of the escaping particles will be overrun by the shock, these particles only fill a much smaller volume near the shock surface than the volume of the particles diffusion region. Therefore, the probability of reacceleration process is quite small.

Our accumulative diffusion model proves successful in explaining the \grays\ from the nine interacting SNRs, which is again indicative
of the contribution of the protons escaping from SNR shocks to the diffusive CRs. This model shows its advantage in treating the hadronic emission which arises from the clouds near the SNR shocks. As a matter of fact, if the distance between the SNR and MC is considerably large than the SNR size, our result will be similar to the point-like injection case.

%perspective CTA,  $\delta\Rc$

\subsection*{Acknowledgments}
We thank the anonymous referee for valuable comments and Stefano Gabici for the helpful advice on diffusion coefficient. Y.C. acknowledges support from NSFC grants 10725312 and the 973 Program grant 2009CB824800.

\bsp
\label{lastpage}

\begin{thebibliography}{99}
\bibitem[Abdo et al.(2009)]{2009ApJ...706L...1A} Abdo, A.~A., et al.\ 2009, \apjl, 706, L1
\bibitem[Abdo et al.(2010)]{2010Sci...327.1103A} Abdo, A.~A., et al.\ 2010a, Science, 327, 1103
\bibitem[Abdo et al.(2010)]{2010ApJ...712..459A} Abdo, A.~A., et al.\ 2010b, \apj, 712, 459
\bibitem[Abdo et al.(2010)]{2010ApJ...718..348A} Abdo, A.~A., et al.\ 2010c, \apj, 718, 348
\bibitem[Abdo et al.(2010)]{2010ApJ...722.1303A} Abdo, A.~A., et al.\ 2010d, \apj, 722, 1303
\bibitem[Acciari et al.(2009)]{2009ApJ...698L.133A} Acciari, V.~A., et al.\ 2009, \apjl, 698, L133
\bibitem[Aharonian \& Atoyan(1996)]{1996A&A...309..917A} Aharonian, F.~A., \& Atoyan, A.~M.\ 1996, \aap, 309, 917
\bibitem[Aharonian(2004)]{2004vhec.book.....A} Aharonian, F.~A.\ 2004, Very high energy cosmic gamma radiation : a crucial window on the extreme
Universe, by F.A.~Aharonian.~River Edge, NJ: World Scientific Publishing, 2004,
\bibitem[Aharonian et al.(2006)]{2006ApJ...636..777A} Aharonian, F., et al.\ 2006, \apj, 636, 777
\bibitem[Aharonian et al.(2008)]{2008A&A...481..401A} Aharonian, F., et al.\ 2008a, \aap, 481, 401
\bibitem[Aharonian et al.(2008)]{2008A&A...490..685A} Aharonian, F., et al.\ 2008b, \aap, 490, 685
\bibitem[Albert et al.(2007)]{2007ApJ...664L..87A} Albert, J., et al.\ 2007, \apjl, 664, L87
\bibitem[Berezinskii et al.(1990)]{1990acr..book.....B} Berezinskii, V.~S., Bulanov, S.~V., Dogiel, V.~A., \& Ptuskin, V.~S.\ 1990, Amsterdam: North-Holland, 1990, edited by Ginzburg, V.L.,
\bibitem[Blandford \& Eichler(1987)]{1987PhR...154....1B} Blandford, R., \& Eichler, D.\ 1987, PhR, 154, 1
\bibitem[Blinnikov et al.(1982)]{1982SvAL....8..361B} Blinnikov, S.~I., Imshennik, V.~S., \& Utrobin, V.~P.\ 1982, Soviet Astronomy Letters, 8, 361
\bibitem[Brogan \& Troland(2001)]{2001ApJ...550..799B} Brogan, C.~L., \& Troland, T.~H.\ 2001, \apj, 550, 799
\bibitem[Brun et al. 2010]{}Brun et al., for the H.E.S.S. collaboration,  Proceedings of the 25th Texas Symposium on Relativistic Astrophysics (Heidelberg, Germany, 2010), to appear in Proceedings of Science
\bibitem[Carmona et al.(2011)]{2011arXiv1110.0950C} Carmona, E., Krause, J., Reichardt, I., \& for the Magic Collaboration 2011, arXiv:1110.0950 
\bibitem[Castelletti et al.(2007)]{2007A&A...471..537C} Castelletti, G., Dubner, G., Brogan, C., \& Kassim, N.~E.\ 2007, \aap, 471, 537
\bibitem[Castro \& Slane(2010)]{2010ApJ...717..372C} Castro, D., \& Slane, P.\ 2010, \apj, 717, 372
\bibitem[Cornett et al.(1977)]{1977A&A....54..889C} Cornett, R.~H., Chin, G., \& Knapp, G.~R.\ 1977, \aap, 54, 889
\bibitem[de Cea del Pozo et al.(2009)]{2009ApJ...698.1054D} de Cea del Pozo, E., Torres, D.~F., \& Rodriguez Marrero, A.~Y.\ 2009, \apj, 698, 1054
\bibitem[Dickman et al.(1992)]{1992ApJ...400..203D} Dickman, R.~L., Snell, R.~L., Ziurys, L.~M., \& Huang, Y.-L.\ 1992, \apj, 400, 203
\bibitem[Dubner et al.(2004)]{2004A&A...426..201D} Dubner, G., Giacani, E., Reynoso, E., \& Par{\'o}n, S.\ 2004, \aap, 426, 201
\bibitem[Fang \& Zhang(2010)]{2010MNRAS.405..462F} Fang, J., \& Zhang, L.\ 2010, \mnras, 405, 462
\bibitem[Fiasson et al.(2008)]{2008AIPC.1085..361F} Fiasson, A., Kosack, K., Skilton, J., Gallant, Y., Hinton, J., P$\ddot{\rm u}$lhofer, G.\ 2008, American Institute of Physics Conference Series, 1085, 361
\bibitem[Frail et al.(1994)]{1994ApJ...424L.111F} Frail, D.~A., Goss, W.~M., \& Slysh, V.~I.\ 1994, \apjl, 424, L111
\bibitem[Frail et al.(1996)]{1996AJ....111.1651F} Frail, D.~A., Goss, W.~M., Reynoso, E.~M., Giacani, E.~B., Green, A.~J., \& Otrupcek, R.\ 1996, \aj, 111, 1651
\bibitem[Fujita et al.(2009)]{2009ApJ...707L.179F} Fujita, Y., Ohira, Y., Tanaka, S.~J., \& Takahara, F.\ 2009, \apjl, 707, L179
\bibitem[Fujita et al.(2010)]{2010ApJ...712L.153F} Fujita, Y., Ohira, Y., \& Takahara, F.\ 2010, \apjl, 712, L153
\bibitem[Fujita et al.(2011)]{2011MNRAS.415.3434F} Fujita, Y., Takahara, F., Ohira, Y., \& Iwasaki, K.\ 2011, \mnras, 415, 3434
\bibitem[Gabici et al.(2009)]{2009MNRAS.396.1629G} Gabici, S., Aharonian, F.~A., \& Casanova, S.\ 2009, \mnras, 396, 1629
\bibitem[Gabici et al.(2010)]{2010sf2a.conf..313G} Gabici, S., Casanova, S., Aharonian, F.~A., \& Rowell, G.\ 2010, SF2A-2010: Proceedings of the Annual meeting of the French Society of Astronomy and Astrophysics, 313
\bibitem[Giuliani et al.(2010)]{2010A&A...516L..11G} Giuliani, A., et al.\ 2010, \aap, 516, L11
\bibitem[Giuliani et al.(2011)]{2011ApJ...742L..30G} Giuliani, A., Cardillo, M., Tavani, M., et al.\ 2011, \apjl, 742, L30 
\bibitem[Graham et al.(1991)]{1991AJ....101..175G} Graham, J.~R., Wright, G.~S., Hester, J.~J., \& Longmore, A.~J.\ 1991a, \aj, 101, 175
\bibitem[Graham et al.(1991)]{1991ApJ...372L..21G} Graham, J.~R., Wright, G.~S., \& Geballe, T.~R.\ 1991b, \apjl, 372, L21
\bibitem[Green(2009)]{2009BASI...37...45G} Green, D.~A.\ 2009, Bulletin of the Astronomical Society of India, 37, 45
\bibitem[Hartman et al.(1999)]{1999ApJS..123...79H} Hartman, R.~C., et al.\ 1999, \apjs, 123, 79
\bibitem[Hewitt et al.(2006)]{2006ApJ...652.1288H} Hewitt, J.~W., Yusef-Zadeh, F., Wardle, M., Roberts, D.~A., \& Kassim, N.~E.\ 2006, \apj, 652, 1288
\bibitem[Hewitt et al.(2009)]{2009ApJ...706L.270H} Hewitt, J.~W., Yusef-Zadeh, F., \& Wardle, M.\ 2009, \apjl, 706, L270
\bibitem[Hoffman et al.(2005)]{2005ApJ...627..803H} Hoffman, I.~M., Goss, W.~M., Brogan, C.~L., \& Claussen, M.~J.\ 2005, \apj, 627, 803
\bibitem[Inoue et al.(2010)]{2010ApJ...723L.108I} Inoue, T., Yamazaki, R., \& Inutsuka, S.-i.\ 2010, \apjl, 723, L108
\bibitem[Jiang et al.(2010)]{2010ApJ...712.1147J} Jiang, B., Chen, Y., Wang, J., Su, Y., Zhou, X., Safi-Harb, S., \& DeLaney, T.\ 2010, \apj, 712, 1147
\bibitem[Kaspi et al.(1993)]{1993ApJ...409L..57K} Kaspi, V.~M., Lyne, A.~G., Manchester, R.~N., Johnston, S., D'Amico, N., \& Shemar, S.~L.\ 1993, \apjl, 409, L57
\bibitem[Kassim et al.(1991)]{1991ApJ...374..212K} Kassim, N.~E., Weiler, K.~W., \& Baum, S.~A.\ 1991, \apj, 374, 212
\bibitem[Katagiri et al.(2011)]{2011arXiv1108.1833K} Katagiri, H., et al.\ 2011, arXiv:1108.1833
\bibitem[Katz \& Waxman(2008)]{2008JCAP...01..018K} Katz, B., \& Waxman, E.\ 2008, JCAP, 1, 18
\bibitem[Kelner et al.(2006)]{2006PhRvD..74c4018K} Kelner, S.~R., Aharonian, F.~A., \& Bugayov, V.~V.\ 2006, \prd, 74, 034018
\bibitem[Keohane et al.(2007)]{2007ApJ...654..938K} Keohane, J.~W., Reach, W.~T., Rho, J., \& Jarrett, T.~H.\ 2007, \apj, 654, 938
\bibitem[Koo \& Moon(1997)]{1997ApJ...475..194K} Koo, B.-C., \& Moon, D.-S.\ 1997a, \apj, 475, 194
\bibitem[Koo \& Moon(1997)]{1997ApJ...485..263K} Koo, B.-C., \& Moon, D.-S.\ 1997b, \apj, 485, 263
\bibitem[Koo et al.(2005)]{2005ApJ...633..946K} Koo, B.-C., Lee, J.-J., Seward, F.~D., \& Moon, D.-S.\ 2005, \apj, 633, 946
\bibitem[Ku et al.(1984)]{1984ApJ...278..615K} Ku, W.~H.-M., Kahn, S.~M., Pisarski, R., \& Long, K.~S.\ 1984, \apj, 278, 615
\bibitem[Levenson et al.(1997)]{1997ApJ...484..304L} Levenson, N.~A., et al.\ 1997, \apj, 484, 304
\bibitem[Levenson et al.(1999)]{1999ApJ...526..874L} Levenson, N.~A., Graham, J.~R., \& Snowden, S.~L.\ 1999, \apj, 526, 874
\bibitem[Li \& Chen(2010)]{2010MNRAS.409L..35L} Li, H., \& Chen, Y.\ 2010, \mnras, 409, L35 (Paper I)
\bibitem[Lozinskaya(1992)]{1992sswi.book.....L} Lozinskaya, T.~A.\ 1992, New York: American Institute of Physics, 1992,
\bibitem[Malkov et al.(2011)]{2011NatCo...2E.194M} Malkov, M.~A., Diamond, P.~H., \& Sagdeev, R.~Z.\ 2011, Nature Communications, 2,
\bibitem[Mehault (2011)]{} M\'ehault, J., Hofverberg, P., Renaudz, M., Cohen-Tanugi, J., Acero, F., Feinstein, F., Grondiny, M.-H., Lemoine-Goumard, M. 2011, talk in the Roma Fermi Symposium
\bibitem[Miyata \& Tsunemi(2001)]{2001ApJ...552..624M} Miyata, E., \& Tsunemi, H.\ 2001, \apj, 552, 624
\bibitem[Moon \& Koo(1994)]{1994JKAS...27...81M} Moon, D.-S., \& Koo, B.-C.\ 1994, Journal of Korean Astronomical Society, 27, 81
\bibitem[Ohira et al.(2011)]{2011MNRAS.410.1577O} Ohira, Y., Murase, K., \& Yamazaki, R.\ 2011, \mnras, 410, 1577
\bibitem[Reach et al.(2006)]{2006AJ....131.1479R} Reach, W.~T., et al.\ 2006, \aj, 131, 1479
\bibitem[Reynoso \& Mangum(2000)]{2000ApJ...545..874R} Reynoso, E.~M., \& Mangum, J.~G.\ 2000, \apj, 545, 874
\bibitem[Reynoso \& Mangum(2001)]{2001AJ....121..347R} Reynoso, E.~M., \& Mangum, J.~G.\ 2001, \aj, 121, 347
\bibitem[Rho \& Petre(1998)]{1998ApJ...503L.167R} Rho, J., \& Petre, R.\ 1998, \apjl, 503, L167
\bibitem[Rosado et al.(2007)]{2007AJ....133...89R} Rosado, M., Arias, L., \& Ambrocio-Cruz, P.\ 2007, \aj, 133, 89
\bibitem[Scoville et al.(1977)]{1977ApJ...216..320S} Scoville, N.~Z., Irvine, W.~M., Wannier, P.~G., \& Predmore, C.~R.\ 1977, \apj, 216, 320
\bibitem[Seta et al.(1998)]{1998ApJ...505..286S} Seta, M., et al.\ 1998, \apj, 505, 286
\bibitem[Seta et al.(2004)]{2004AJ....127.1098S} Seta, M., Hasegawa, T., Sakamoto, S., Oka, T., Sawada, T., Inutsuka, S.-i., Koyama, H., \& Hayashi, M.\ 2004, \aj, 127, 1098
\bibitem[Sezer et al.(2011)]{2011arXiv1107.1054S} Sezer, A., G{\"o}k, F., Hudaverdi, M., \& Ercan, E.~N.\ 2011, arXiv:1107.1054
\bibitem[Slane et al.(2002)]{2002ApJ...580..904S} Slane, P., Chen, Y., Lazendic, J.~S., \& Hughes, J.~P.\ 2002, \apj, 580, 904
\bibitem[Strong et al.(2007)]{2007ARNPS..57..285S} Strong, A.~W., Moskalenko, I.~V., \& Ptuskin, V.~S.\ 2007, Annual Review of Nuclear and Particle Science, 57, 285
\bibitem[Tang et al.(2011)]{2011ApJ...739...11T} Tang, Y.~Y., Fang, J., \& Zhang, L.\ 2011, \apj, 739, 11
\bibitem[Tavani et al.(2010)]{2010ApJ...710L.151T} Tavani, M., et al.\ 2010, \apjl, 710, L151
\bibitem[Tian et al.(2007)]{2007ApJ...657L..25T} Tian, W.~W., Li, Z., Leahy, D.~A., \& Wang, Q.~D.\ 2007, \apjl, 657, L25
\bibitem[Torres et al.(2008)]{2008MNRAS.387L..59T} Torres, D.~F., Rodriguez Marrero, A.~Y., \& de Cea Del Pozo, E.\ 2008, \mnras, 387, L59
\bibitem[Torres et al.(2010)]{2010MNRAS.408.1257T} Torres, D.~F., Marrero, A.~Y.~R., \& de Cea Del Pozo, E.\ 2010, \mnras, 408, 1257
\bibitem[Torres et al.(2011)]{2011arXiv1107.3470T} Torres, D.~F., Li, H., Chen, Y., et al.\ 2011, arXiv:1107.3470
\bibitem[Uchiyama et al.(2010)]{2010ApJ...723L.122U} Uchiyama, Y., Blandford, R.~D., Funk, S., Tajima, H., \& Tanaka, T.\ 2010, \apjl, 723, L122
\bibitem[Vel{\'a}zquez et al.(2002)]{2002AJ....124.2145V} Vel{\'a}zquez, P.~F., Dubner, G.~M., Goss, W.~M., \& Green, A.~J.\ 2002, \aj, 124, 2145
\bibitem[Yamazaki et al.(2006)]{2006MNRAS.371.1975Y} Yamazaki, R., Kohri, K., Bamba, A., Yoshida, T., Tsuribe, T., \& Takahara, F.\ 2006, \mnras, 371, 1975
\bibitem[Zhang \& Fang(2008)]{2008ApJ...675L..21Z} Zhang, L., \& Fang, J.\ 2008, \apjl, 675, L21
\bibitem[Zhou et al.(2010)]{2010MNRAS.406..223Z} Zhou, X., Bocchino, F., Miceli, M., Orlando, S., \& Chen, Y.\ 2010, \mnras, 406, 223

\end{thebibliography}
\end{document}